\documentclass[twocolumn]{aastex631}
\usepackage{amsmath}
\usepackage{xcolor}
\definecolor{forestgreen(web)}{rgb}{0.13, 0.55, 0.13}

\usepackage{xcolor}
\begin{document}

\title{Localized Heating and Dynamics of the Solar Corona due to a Symbiosis of Waves and Reconnection}

\author{A.K.~Srivastava}
\affiliation{Department of Physics, Indian Institute of Technology (BHU), Varanasi-221005, India. Email:- asrivastava.app@
iitbhu.ac.in}
\author{Sripan Mondal}
\affiliation{Department of Physics, Indian Institute of Technology (BHU), Varanasi-221005, India}
\author{Eric R. Priest}
\affiliation{Mathematics Institute, St Andrews University, KY16 9SS, St Andrews, UK}
\author{Sudheer K. Mishra}
\affiliation{Astronomical Observatory, Kyoto University, Sakyo, Kyoto 606-8502, Japan}
\author{David I. Pontin}
\affiliation{School of Information and Physical Sciences, University of Newcastle, Australia}
\author{R.Y.~Kwon}
\affiliation{Korea Astronomy and Space Science Institute, Daejeon 34055, Republic of Korea}
\author{Ding Yuan}
\affiliation{Shenzhen Key Laboratory of Numerical Prediction for Space Storm, Institute of Space Science and Applied Technology, Harbin Institute of Technology, Shenzhen, Guangdong, People's Republic of China}
\author{K.~Murawski}
\affiliation{Institute of Physics, University of Maria Curie-Sklodowska, Pl. M. Curie-Sklodowskiej 5, 20-031, Lublin, Poland}
\author{Ayumi~Asai}
\affiliation{Astronomical Observatory, Kyoto University, Sakyo, Kyoto 606-8502, Japan}

\begin{abstract}
The Sun’s outer atmosphere, the corona, is maintained at mega-Kelvin temperatures and fills the heliosphere with a supersonic outflowing wind. The dissipation of magnetic waves and direct electric currents are likely to be the most significant processes for heating the corona, but a lively debate exists on their relative roles. Here, we suggest that the two are often intrinsically linked, since magnetic waves may trigger current dissipation, and impulsive reconnection can launch magnetic waves. We present a study of the first of these processes 
by using a 2D physics-based numerical simulation using the Adaptive Mesh Refined (AMR) Versatile Advection Code (VAC). 
Magnetic waves such as fast magnetoacoustic waves are often observed to propagate in the large-scale corona and interact with local magnetic structures.
The present numerical simulations show how the propagation of magnetic disturbances towards a null point or separator can lead to the accumulation of the electric currents. Lorentz forces can laterally push and vertically stretch the magnetic fields, forming a current sheet with a strong magnetic-field gradient. The magnetic field lines then break and reconnect, and so contribute towards coronal heating. 
Numerical results are presented that support these ideas and support the concept of a symbiosis between waves and reconnection in heating the solar corona. 
\end{abstract}


\section{Introduction}

The corona, the outermost layer of the solar atmosphere, is subject to energy and mass loss due to thermal conduction, supersonic solar wind outflows, and radiative emissions. However, it always maintains a high temperature regardless of those energy losses.
In addition, sufficient mass must be available at coronal heights to supply the solar wind. 
Therefore, the sustained mega-Kelvin temperature in the corona, the origin of mass transport in the nascent solar wind, and associated various transient plasma processes are key problems in solar physics and plasma astrophysics that are widely debated \citep[e.g., see][]
{2014masu.book.....P}. The damping of magnetohydrodynamic (MHD) waves and the dissipation of current via magnetic reconnection are likely to contribute to the heating, mass transport, and other transient processes in the Sun's atmosphere \citep[e.g., see][]{2005LRSP....2....3N}. Resonant absorption \citep[e.g.,][]{2004prma.book.....G,2011SSRv..158..289G}
and phase mixing \citep[e.g.,][]{1983A&A...117..220H,1998A&A...338.1118R,1999A&A...346..641D,2020A&A...636A..40H}
have been proposed as dissipation mechanisms through which particular waves may dissipate and provide heat for the corona \citep[e.g.,][and references cited therein]{2020SSRv..216..140V}. On the other hand, heating can also take place during magnetic reconnection in braided magnetic fields \citep[e.g.,][]{
2020LRSP...17....5P}, kink-unstable twisted fields \citep[e.g.,][]{2016ApJ...817....5H},
due to coronal tectonics driven by the magnetic carpet \citep[e.g.,][]{2002ApJ...576..533P}, as well as in the magnetic nulls and separators \citep[e.g.,][]{2022LRSP...19....1P}.

Various MHD wave modes can be excited at the base of the photosphere due to the presence of turbulent convection subjected to different types of motions at their footpoints, which further propagate upward through various magnetic field structures \citep[e.g.,][and references cited therein]{2019mwsa.book.....R,2020ARA&A..58..441N}.
\citet{2011ApJ...727...17F} 
reported that high-frequency magnetoacoustic waves can leak energy effectively into the corona. Torsional Alfv\'en waves are also established to carry enough energy to the coronal heights even after their partial reflection from the solar transition region \citep[e.g.,][]{2017NatSR...743147S,2019ApJ...871....3S}.
In addition, there can be {\it in situ} generation and evolution of different wave modes in the solar corona itself through magnetic reconnection during flares, coronal mass ejections, and filament eruptions \citep[e.g.,][and references cited therein]{2012RSPTA.370.3193D}.
Imaging and spectroscopic observations using space-borne observatories such as SOHO/EIT 
, STEREO/EUVI 
, HINODE/EIS 
, SDO/AIA 
, etc along with ground-based observations using CoMP \citep[e.g.,][]{2008SoPh..247..411T}
have established
the properties of MHD waves that are ubiquitously present in the solar corona and have characterized them in terms of slow, fast magnetoacoustic, Alfv\'en waves, and other tubular modes such as kink, sausage, torsional Alfv\'en waves. Due to multi-wavelength observations via Hinode, IRIS \citep{2014SoPh..289.2733D}, AIA, and other coordinated observations using space- and ground-based observatories, different wave processes such as reflection, refraction, mode conversion,  
and dissipation have been captured by detecting different atmospheric layers of the Sun simultaneously.
In particular, the observational manifestation of fast magnetoacoustic waves and Alfv\'en waves in the large-scale solar atmosphere, generated by the magnetic reconnection and flaring processes in the corona, has been studied extensively.  

Magnetic reconnection, a topological or geometrical reconstruction of the magnetic field coupled to the plasma, is important because it can convert stored magnetic energy into various forms such as heat, bulk kinetic energy and fast particle energy. Moreover, it may be part of the underlying mechanism that accelerates the suersonic solar wind and causes energetic flares and coronal mass ejections, which in turn have large-scale impacts in the role of space weather events \citep[e.g.,][and references cited therein]{2014masu.book.....P,2022LRSP...19....1P,2023ApJ...945...28R}.

In 2-D descriptions, magnetic X-points are the only locations where magnetic reconnection can take place. But in reality, reconnection is a 3-D process that can take place at magnetic nulls, separators,  and  quasi-separators, as well as by braiding \citep[e.g.,][]{2022LRSP...19....1P}. The presence of null points is inferred throughout the solar chromosphere and corona from magnetic field extrapolations \citep[e.g.,][]{2009SoPh..254...51L}. But, irrespective of a region possessing a magnetic null, separator or quasi-separator, it is the development of a region with  high current density around such a feature, i.e., a current sheet with an abrupt gradient in the magnetic field, that is essential for magnetic reconnection. The formation of a current sheet and the triggering of magnetic reconnection can be initiated via a localized enhancement of resistivity  
or via an external perturbation (i.e., by externally driven or forced reconnection).
 Even though forced reconnection has long been studied theoretically, \citet{2019ApJ...887..137S} gave an example of the forced  reconnection at a null point in the solar corona, where the current sheet had formed due to an inflow created by a distant prominence eruption. This was the first direct imaging observation of the forced magnetic reconnection. To follow up, \citet{2021ApJ...920...18S} studied an example of the generation of coronal jet-like features and hot plasma flows in the off-limb solar corona. Recently, \citet{2025arXiv250212889M} have observationally studied the jet-driven evolution of the forced magnetic reconnection region in a prominence segment where bi-directional outflows, heating, and plasmoids were evident.

Thus, both magnetic reconnection and different MHD wave modes are ubiquitous in the solar corona, and they may co-exist in the quiet-Sun (QS) as well as in active regions (ARs). The effects of these two processes are enhanced by their co-existence and mutual interaction, which has been referred to as a ``symbiosis"  \citep{Sri24,2024ApJ...977..235M}.
The relative contribution of waves and reconnection in heating the different parts of the solar atmosphere has not yet been determined. 
Therefore, it will be significant to understand the role of a symbiosis between the two in localized coronal heating and plasma dynamics. There have been several studies on the interaction of MHD waves with magnetic nulls in coronal conditions and gravitationally stratified atmospheres both observationally and theoretically \citep[e.g.,][and references cited therein]{2017ApJ...834L..15Z,2022MNRAS.515.5094G,2024NatCo..15.2667K}.
Typical physical processes such as mode conversion, current sheet formation and onset of reconnection are reported as outcomes of such interactions.
However, there was relatively little direct observational evidence of these interactions until   mode conversion of MHD waves near a 3D magnetic null and coronal cavity was reported \citep[e.g.,][and references cited therein]{2017ApJ...834L..15Z,2018ApJ...863..101C,2024NatCo..15.2667K}.  These reports, however, could not demonstrate observations of current sheet formation and reconnection onset, with subsequent heating and plasma dynamics. A direct heating scenario was also not detected during such a mode conversion \citep{2024NatCo..15.2667K}. 

The proposed symbiosis depends on the co-existence of waves and reconnection, which help to reinforce each other during their roles in heating and  plasma dynamics. 
What we are calling SWAR (a Symbiosis of WAves and Reconnection)  represents a “mutualism” between waves and reconnection in  solar  plasmas \citep[e.g.,][]{Sri24,2024ApJ...977..235M}. These two processes may drive each other, and so together play a role in causing localized heating and plasma dynamics in the solar atmosphere.
There are several reports depicting how magnetic reconnection initiates waves \& oscillations or vice-versa in the context of the solar plasma. However, SWAR presents an overview of their origin, likelihood in space and time, physical co-existence (dependence of one's properties on the other), and mutualism (reinforcement and support) \citep[see,][]{Sri24,2024ApJ...977..235M}.

Fragmentation and reconnection in current sheets is known to drive MHD waves and oscillations and vice-versa \citep[e.g.,][and references cited therein]{2007PhPl...14l2905L,2018ApJ...868L..33L,2024ApJ...977..235M}.
Also, plasmoid coalescence has been seen to drive waves from a current sheet \citep[e.g.,][references cited there]{2015ApJ...800..111Y,2016ApJ...823..150T,2024ApJ...977..235M}.
From an observational point of view, various literature also demonstrate that waves from a flaring region may encounter a coronal null, which may then drive reconnection in various regimes \citep[e.g.,][and references cited there]{2017ApJ...834L..15Z,2018ApJ...863..101C,2024NatCo..15.2667K}. 
In the present paper, we describe one aspect using a numerical model of wave-driven collapse of a null region and localized heating and dynamics of the solar corona. In Section 2, we present the numerical model, namely, the thinning process and localized heating in a reconnection region initiated by a wave-like perturbation.  In Section 3, we elaborate on the possible role of SWAR as a viable physical mechanism for coronal heating and plasma dynamics in the solar corona. Finally, we present the conclusions and future aspects of the work.

\section{A Numerical Example of SWAR}
We aim here to demonstrate the underlying physics of the formation of a current-sheet-like structure in the corona and the onset of reconnection after the passage of a 
wave-like disturbances using a 2-D numerical MHD simulation.
\subsection{A Description of the MHD Model}
 The following set of resistive, conservative magnetohydrodynamic (MHD) equations \citep{2014masu.book.....P,2017ApJ...841..106Z,2024ApJ...977..235M} is solved numerically using open source MPI-AMRVAC 3.0\footnote{https://amrvac.org/} \citep{2023A&A...673A..66K}:
\begin{equation}
\frac{\partial \rho}{\partial t} + \vec{\nabla} \cdot ( \rho \vec{V} ) = 0
\end{equation}
\begin{equation}
  \frac{\partial}{\partial t}(\rho \vec{V}) + \vec{\nabla} \cdot \left [ \rho \vec{V}\vec{V}  + p_{tot}\vec{I} - \frac{\vec{B}\vec{B}}{4\pi} \right ] = 0,
\end{equation}

\begin{equation}
\begin{split}
\frac{\partial e}{\partial t} +  \vec{\nabla} \cdot \left( e\vec{V} + p_{tot}\vec{V} -\frac{\vec{B}\vec{B}}{4\pi} \cdot \vec{V}\right)  = \eta \vec{J^{2}}-
   \vec{B} \cdot \vec{\nabla} \times (\eta \vec{J})\\
   +\vec{\nabla}_{\parallel} \cdot (\kappa_{\parallel} \vec{\nabla}_{\parallel} T),
\end{split}   
\end{equation}

\begin{equation}
  \frac{\partial \vec{B}}{\partial t} + \vec{\nabla} \cdot \left(\vec{V}\vec{B} - \vec{B}\vec{V}\right)+ \vec{\nabla} \times (\eta \vec{J}) = 0,
\end{equation}

\quad \textrm{where} \quad
\begin{equation}
p_{tot} = p + \frac{B^2}{8\pi}, ~~e = \frac{p}{\gamma-1} + \frac{1}{2}\rho V^{2} + \frac{B^2}{8\pi}
\end{equation}
\quad \textrm{and} \quad
\begin{equation}
  \vec{J} = \frac{\vec{\nabla} \times \vec{B}}{4\pi}, ~~\vec{\nabla} \cdot \vec{B} =0.
\end{equation} 
In the initial state, mass density ($\rho$) and temperature (T) are taken to be $2.34 \times 10^{-15}~\mathrm{g~cm^{-3}}$ and 1 MK, respectively, throughout the simulation domain to agree with  typical solar coronal conditions \citep{
2024ApJ...963..139M}. Using the ideal gas law, we estimate the plasma pressure ($p$) to be $ 0.32~\mathrm{dyne~cm^{-2}}$. We adopt a uniform magnetic diffusivity $\eta = 6\times 10^{8}~ \mathrm{m^{2}s^{-1}}$, which is larger than typical coronal values. $\kappa_{\parallel}=10^{-6}~T^{5/2}~\mathrm{erg~cm^{-1}~s^{-1}~K^{-1}}$ is the component of the thermal conduction tensor parallel to the magnetic field. For simplicity, gravity and radiative cooling are not taken into account in this work.

We consider a potential magnetic field as our initial state, namely \citep{2014masu.book.....P,2024ApJ...977..235M}: 
\begin{equation}
f(x,y)= \frac{D[(x-a_{1})^2-(y-a_{2})^2]}{[(x-a_{1})^2-(y-a_{2})^2]^2 + 4(x-a_{1})^2(y-a_{2})^2}, \nonumber\\
\end{equation}
\begin{equation}
B_{x}(x,y) = b_{0} + f(x,y);
\end{equation}
\begin{equation}
B_{y}(x,y)=\frac{2D(x-a_{1})(y-a_{2})}{[(x-a_{1})^2-(y-a_{2})^2]^2 + 4(x-a_{1})^2(y-a_{2})^2};
\end{equation}
and 
\begin{equation}
B_{z}(x,y)=0;
\end{equation}
where $b_{0}=16~\mathrm{Gauss}$, $a_{1}= 0~\mathrm{Mm}$, $a_{2}= -60~\mathrm{Mm}$ and $D= 3.6\times 10^{5}~\mathrm{Gauss~Mm^{2}}$. This configuration results in an average magnetic field of 20 Gauss in the simulation domain with a maximum of 82 Gauss just above the bottom boundary typical of an active region and its  surroundings \citep[e.g.,][]{2020ApJ...898L..34S}. The 2D unstratified simulation domain is in magnetohydrostatic equilibrium initially with $\vec{J} = \vec{0}$ and  contains  a magnetic null 
point  at
$ x = 0~\mathrm{Mm}$, $y = 92~\mathrm{Mm}$. A high-$\beta$ region around the null lies almost in the middle of the simulation box. There is no significant effect of reflected disturbances in the formation of the current sheet due to the collapse of the magnetic null. We have also confirmed that, in the absence of an external perturbation, this magnetic configuration with a magnetic null remains in equilibrium and does not collapse.

\subsection{Velocity Perturbations Generating a Fast Magnetoacoustic Wave}
\begin{figure*}
\hspace{-1.5 cm}
\includegraphics[scale=0.8]{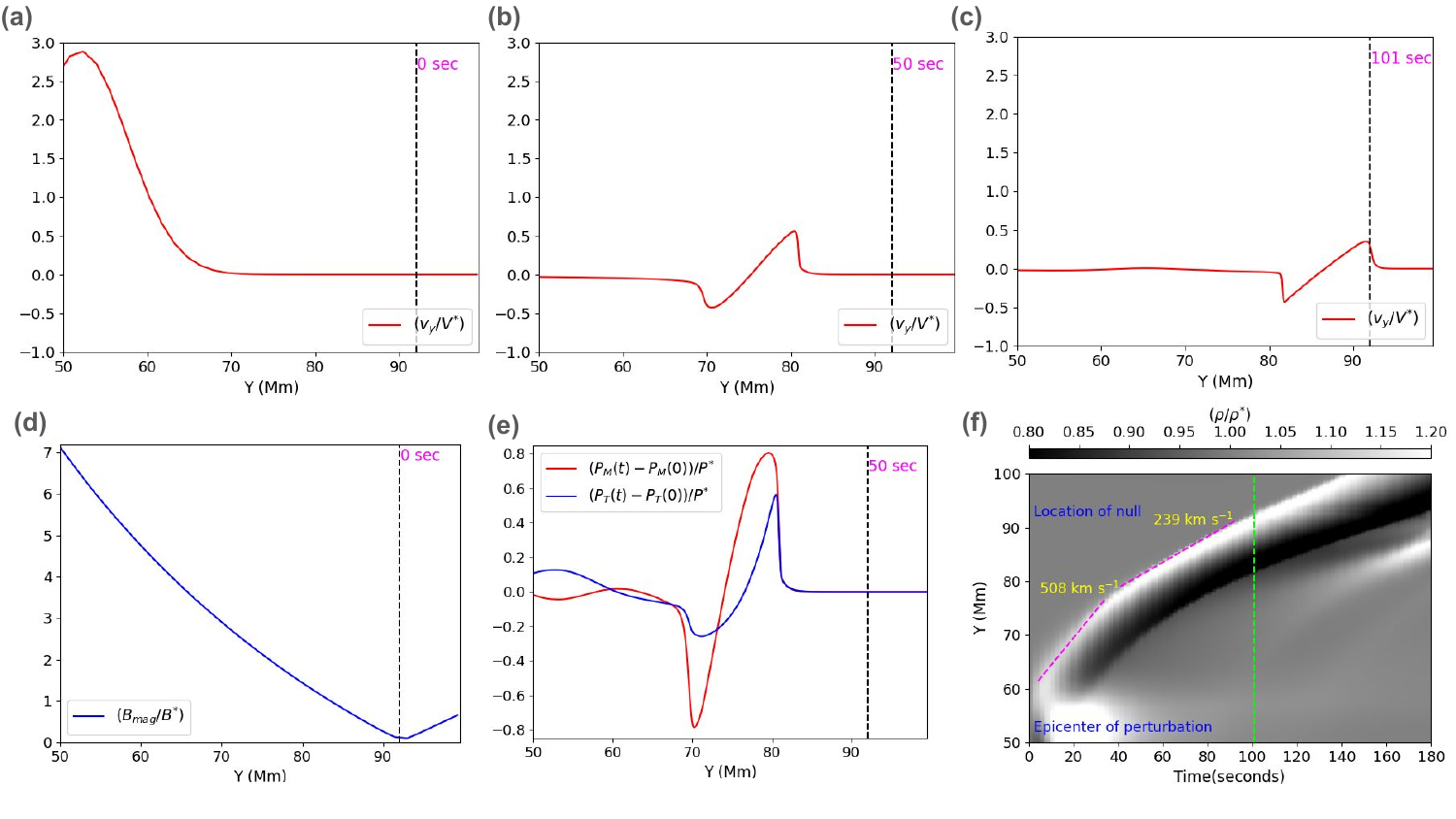}
\caption{Panel (a), (b) and (c) represent the evolution of an initial Gaussian velocity disturbance during its passage towards a magnetic null placed at Y = 92 Mm (denoted via black dashed vertical line). At 0 s, the disturbance is fully Gaussian with a peak at Y = 52 Mm. As time progresses, the disturbance becomes distorted to form a leading sawtooth profile that indicates the formation of a magnetoacoustic shock at 50 s. At about 100 s, the shock wave of amplitude around $50~\mathrm{km.s^{-1}}$ interacts with the magnetic null. Panel (d)  displays a gradual decrease in the magnitude of ambient magnetic field from Y = 52 Mm to Y = 92 Mm, which leads to a deceleration of the leading edge to form the sawtooth profile. Panel (e) shows that the perturbations in normalized  magnetic (red curve) and plasma pressure (blue curve) in the wake of the shock-like features in Panel (b) possess an in-phase relation, which indicates that the shock is fast-mode in nature. Panel (f) is a distance-time diagram showing the evolution in density as the velocity disturbance propagates. As expected, the disturbance  propagates away from its origin at $508~\mathrm{km.s^{-1}}$ and later decelerates to $239~\mathrm{km.s^{-1}}$ while reaching towards the null.}
\label{label 1a}
\end{figure*}

We impose a transient velocity pulse directed upward vertically using a Gaussian profile \citep{2024ApJ...963..139M,2024ApJ...977..235M} at the initial time:
\begin{equation}
   V_{y} = V_{0}~\mathrm{exp} \left(- \frac{(x-x_{0})^{2}}{w_{x}^{2}}-\frac{(y-y_{0})^{2}}{w_{y}^{2}}\right),
\end{equation}
where $x_{0}= 0~\mathrm{Mm}$, $y_{0}= 52~\mathrm{Mm}$, $w_{x}= 4~\mathrm{Mm}$, $w_{y}= 8~\mathrm{Mm}$, and $V_{0} = 350 ~\mathrm{km\ s^{-1}}$  (see Figure~\ref{label 1a}(a)). This velocity pulse produces a fast propagating disturbance that perturbs the vicinity of the null point in a manner that mimics a fast EUV wave propagating towards it and impinged. 
The spatio-temporal evolution and physical characteristics of the leading edge of the imposed velocity perturbation during its passage towards the magnetic null is described below. 

We are interested in the properties of the velocity disturbance just before or at the time of its interaction with the magnetic null, as shown in Figure~\ref{label 1a}(b), which indicates a deformation of the initial Gaussian shape to form a sawtooth profile and so the formation of magnetoacoustic shock waves around 50 s. The amplitude decreases to around $0.5V^{*}$, i.e., $58~\mathrm{km\ s^{-1}}$. Eventually, the forward-moving shock interacts with the magnetic null (as depicted via vertical dashed black lines in panel (a), (b) and (c) of Figure~\ref{label 1a}) around 100 s, with an even smaller amplitude of roughly $0.35V^{*}$, i.e., $40~\mathrm{km\ s^{-1}}$. The formation of the shock and the gradual decrease in amplitude of the disturbance during its propagation are due to the nonlinear nature of the disturbance. Figure~\ref{label 1a}(d) indicates the physical reasons behind the formation of the shock, showing that the magnitude of the ambient magnetic field decreases towards the null.
This leads to a lower local Alfv\'en speed as the disturbance moves forward towards the null, which in turn results in the steepening of the leading edge of the disturbance to form the sawtooth profile. Also, the in-phase relation of thermal and magnetic pressure in the wake of the forward moving shock suggests that the shock is basically fast-mode in nature (see Figure~\ref{label 1a}(e)). Furthermore, a time-distance diagram in density shows the spatio-temporal evolution of the initial disturbance right from its initial position. The diagram provides further evidence of a propagating magnetoacoustic shock wave like structure in the form of higher but opposite density gradients. Since the magnetic field is decreasing towards the magnetic null as shown in Figure~\ref{label 1a}{d}, the disturbance undergoes a deceleration from an initial speed of around $508~\mathrm{km\ s^{-1}}$ to $239~\mathrm{km\ s^{-1}}$ before reaching the magnetic null. Therefore, the formation of shock, a decrement in velocity amplitude and a deceleration of its propagation speed confirm the non-linear nature of the velocity disturbance. All of these properties along with the in-phase relation of thermal and magnetic pressure hold true for non-linear fast magnetosonic waves confirm that our choice of velocity disturbance mimics the interaction of such a wave with the magnetic null.

Our aim is to put forward a physical model that depicts
the collapse of a coronal null to form a current sheet due to the incidence and passage of fast magnetoacoustic waves.  The choice of 350$~\mathrm{km\ s^{-1}}$ as the initial velocity amplitude is high in order to produce a shock wave, since the observed  EUV waves are seen to interact with various coronal structures \citep[e.g.,][]{2013ApJ...766...55K} and steepen due to non-linear effects into shocks before  subsequently decaying during their further evolution in the large-scale corona \citep[e.g.,][]{2023A&A...676A.144M}. Whether  a shock profile is produced or not depends on the initial amplitude as well as the ambient conditions of the plasma and magnetic field.  On reaching the null region, it already diminishes significantly to a few tens of kilometers per second. The phase speed of the fast wave is a typical value depending on local values of the density, magnetic field, and temperature.
During observation of EUV waves propagating as fast magnetoacoustic shocks, \citet{2023A&A...676A.144M} found reasonably larger amplitude of the wave far from its source region compared to the one obtained in our simulation during its interaction with the magnetic null. Therefore, our choice of initial high amplitude of velocity perturbations seem physically reasonable in the framework of EUV waves.

We find similar dynamical processes for lower values of the initial amplitude of the velocity pulse (namely, one tenth of the amplitude, i.e., 35$~\mathrm{km\ s^{-1}}$). The low-amplitude perturbations generate propagating fast magnetoacoustic waves (EUV waves) without steepening into shocks. They also decelerate when reaching the null region. The wave-like perturbations cause  the null region to collapse, but this occurs at a much later time, followed by  current-sheet formation, plasma flows, heating, and plasmoid formation. Therefore, the basic physics of SWAR remains valid, and only the time-line  changes significantly.

\subsection{Numerical Methods}
The spatial range of the simulation domain is $x=~[-100, 100]$~Mm and $y=[0, 200]$~Mm. In both directions, the primary spatial resolution is 1.56 Mm which falls to 48 km after five levels of adaptive mesh refinement (AMR), equivalent to [4096, 4096] grid points.  MPI-AMRVAC works in dimensionless physical variables, such that the physical quantities in Eqs. (1-10) are normalized with respect to typical physical values, namely, length $L^{*}= 10~\mathrm{Mm}$, magnetic field $B^{*}= 2~\mathrm{Gauss}$ and velocity $V^{*}= 116.45~\mathrm{km~s^{-1}}$ \citep{2024ApJ...963..139M}. 
We use a $``$two-step$"$ method for temporal integration and a $``$Harten-Lax-van Leer (HLL)$"$ Riemann solver 
to estimate the flux at cell interfaces. Continuous boundary conditions are used at all the boundaries, so that the gradient of every variable remains zero across them. A second-order symmetric TVD limiter $``$vanleer$"$ 
is employed to suppress spurious numerical oscillations.
\begin{figure*}
\hspace{-1.9 cm}
\includegraphics[scale=0.85]{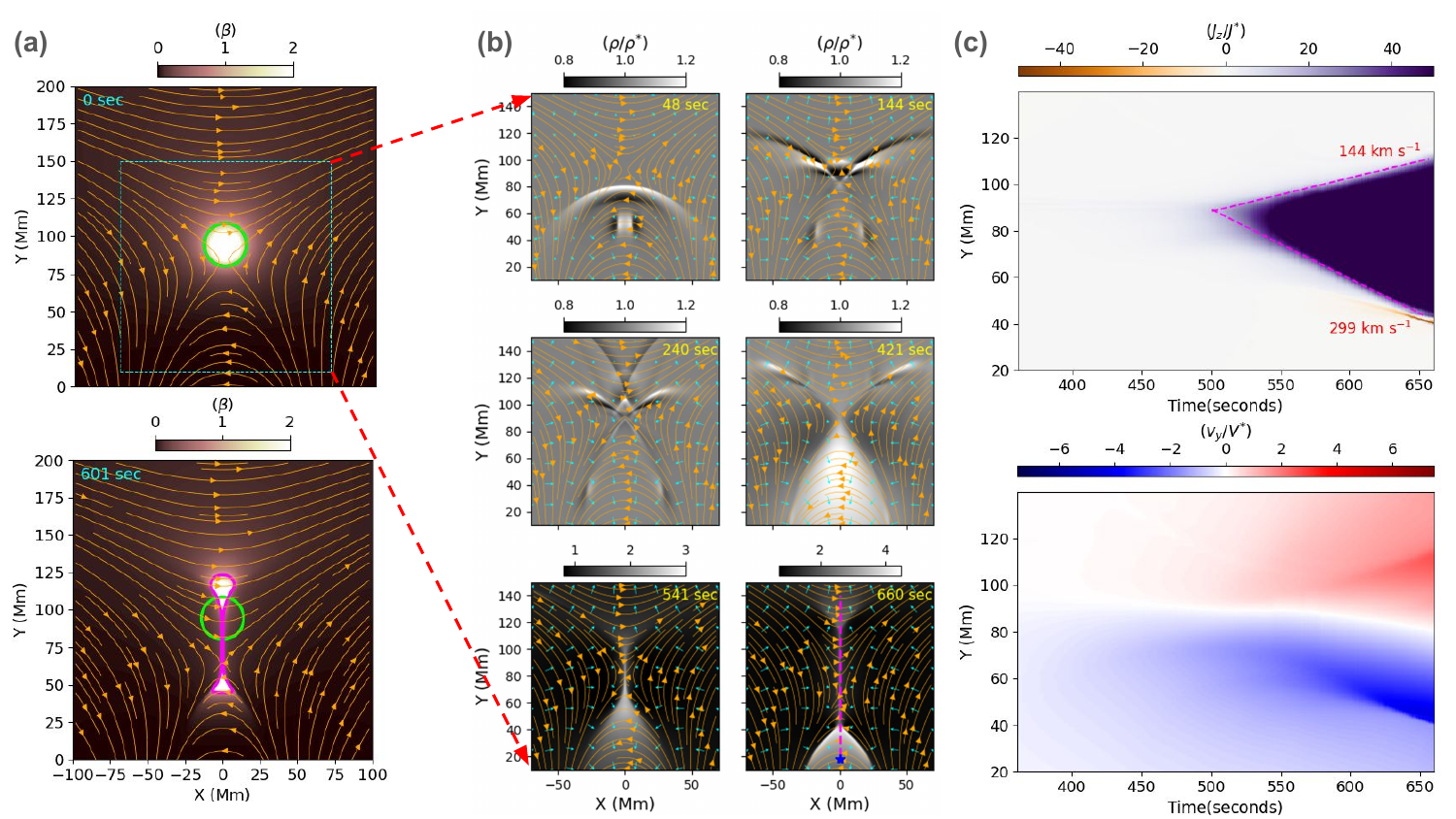}
\caption{(a) Initial plasma $\beta$ distribution at 0 s (top panel) and the time-evolved collapsed high $\beta$ region at 601 s (bottom panel). The lime-coloured curve depicts the initial plasma $\beta = 1$ contour in both the panels, whereas the magenta curve reveals the distorted plasma $\beta = 1$ region.
(b) The propagation of fast magnetoacoustic perturbations, and  their interaction with the coronal magnetic field. After the interaction of part of the wave and its passage through the null region, a current sheet forms. The magenta dashed line starting from the blue star in the bottom right sub panel shows the location of the slit used to extract distance-time diagrams shown in panel (c). (c) The upper panel reveals the gradual enhancement of current density and an elongation of the current enhanced region over time along the slit. The lower panel suggests the presence of bi-directional outflows, which describes the presence of reconnection within the current-enhanced region. 
An animation of the entire process is provided in Figure2.mp4 which runs with a real-time duration of 27-s equivalent to a physical time of 660 s.}
\label{label 2a}
\end{figure*}
\begin{figure*}
\hspace{-1.95cm}
\includegraphics[width=1.225\textwidth]{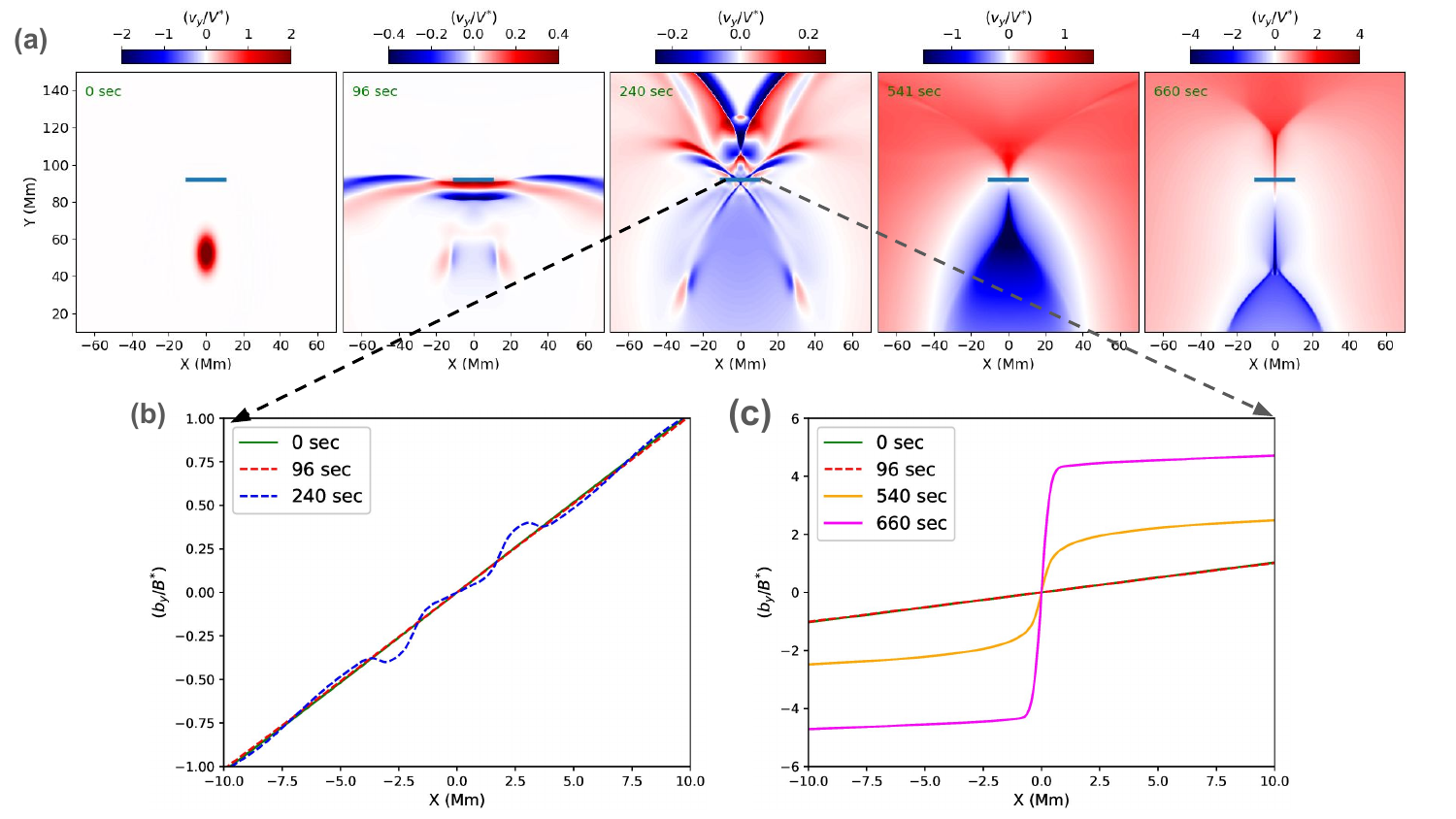}
\caption{Formation of the current sheet by the incidence of waves on the null region: Panel (a) shows the initial velocity pulse at $t$=0 sec, the wavefront at $t$=96 sec just before reaching the height at which the perturbed magnetic field is measured (shown as a horizontal thick cyan line), the other maps of the $y$-component of velocity beyond 240 seconds after the wave-like perturbations have passed through the high \(\beta\) region. The bi-directional outflows at two later times ($t$=541 and 660 s) are consistent with reconnection in the current sheet. Panel (b) describes the change in inflow magnetic field close to the magnetic null due to the passage of the wave through the region (blue dashed line). Panel (c) exhibits a similar physical scenario, but it shows a gradual thinning of the current sheet via steepening of the magnetic field with time. 
}
\label{label 3a}
\end{figure*}
\section{Results}
The results are described below, including a description of how a current sheet and its accompanying reconnection are initiated by a wave.

In Figure~\ref{label 2a} an overview of the dynamics observed in the simulations is presented. Figure~\ref{label 2a}(a) depicts the evolution of a magnetic null and its environment  from their initial configuration (top panel) to a deformed one at later stage, i.e., at 601 s (bottom panel).
Initially, the wavefront propagates with a dome-shaped appearance (see Figure~\ref{label 2a}(b), $t$=48 s panel, indicating an approximately spatially-uniform wave speed, before becoming deformed like a bird's wing once it 
reaches the high-$\beta$ region around the null point,
as a result of the trapping of the middle portion of the wavefront (see Figure~\ref{label 2a}(b), $t=144$~s panel). The rest of the wave-front continues to propagate in the low plasma-$\beta$ region, without deformation. 
 Secondary perturbations that move outward from the null region are detected, while counter-propagating slow perturbations are also seen in the low-lying loop-like structures connected to the $y=0$ boundary (see Figure~\ref{label 2a}(b), $t=240$~s onward). Although, our main focus in the present work is to understand the formation, dynamics, and heating of the current sheet.
  
In our simulation, where the wave energy is larger than considered by \citet{2024NatCo..15.2667K}, the entire localized coronal structure becomes vertically stretched and laterally thins in time,
forming
a current sheet (CS), favourable for the onset of reconnection, heating, and associated plasma dynamics (see Figure~\ref{label 2a}(b), $t=541$~s onward). Figure~\ref{label 2a}(c) (top-panel) demonstrates gradual enhancement and elongation of the current density along the vertical slit depicted via the magenta dashed vertical line in the 660 sec sub-panel of panel (b). 
Figure~\ref{label 2a}(c) (bottom panel) shows the generation of bi-direction flows at a later time once the current sheet is formed.  Together this illustrates the formation of a reconnecting current sheet by a propagating fast magnetoacoustic wave perturbations. 
The entire dynamics in the numerical simulation is also shown in the animation Figure2.mp4 associated with Figure~\ref{label 2a}.  

\begin{figure*}
\hspace{-2.2 cm}
\includegraphics[scale=0.9]{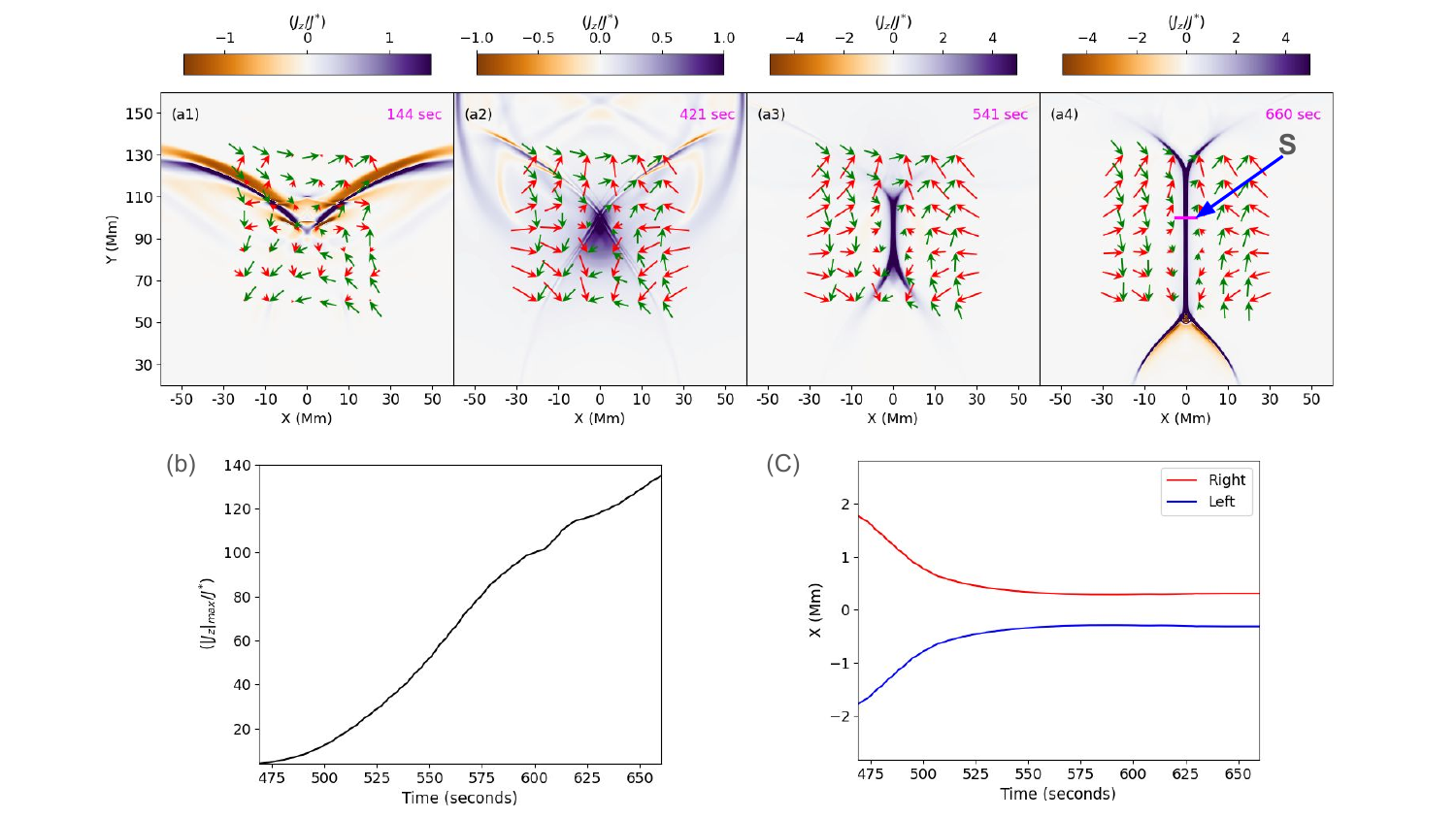}
\caption{The effect of the Lorentz force on the formation of a current sheet: The top-row shows the normalized current-density maps overlaid by  magnetic field  (green arrows) and  Lorentz force vectors (red arrows).  
The wave passes through the corona, and a portion of its wave-front  
 perturbs the magnetic field and creates localized currents. The currents and associated Lorentz force distributed around the null help to form the current sheet.
The Lorentz force stretches and thins the magnetic field to form the current sheet and triggers magnetic reconnection. 
(b) The enhancement in the maximum of the current density with time measured at Y = 92 Mm, i.e., the initial location of the magnetic null, which indicates a gradual accumulation of current at that location from 469 s to 660 s. (c) shows the gradual thinning of the current sheet starting from 469 s to 660 s. At 469 s the current density profile can be fitted with a Gaussian profile to estimate the FWHM as a quantitative measure of the CS width.  
The accompanying animation, Figure4.mp4, demonstrates the effect of the Lorentz force on the coronal null region, the formation of the current sheet, and the trigger of reconnection, which for a real-time duration of 27-sec equivalent to a physical duration of 660 sec.
}
\label{label 4a}
\end{figure*}

The initial state of the present numerical simulation is in  quasi-static equilibrium with a \textbf{current-free}  magnetic field. 
As shown in Figure~\ref{label 3a}, prior to the arrival of the fast wave the magnetic field around the null is unperturbed.
Figure~\ref{label 3a} demonstrates that, before the wave impinges upon the null region ($t$ = 0 s in (a)), the magnetic field is unperturbed there (see panel (b)). Similarly, at $t$ = 96 s,  the wave is about to reach the cyan horizontal slit. At this time,
no signature of the perturbation of the magnetic field is evident (see panel (b) and (c)).
However, when part of the wave passes through the null region and perturbs it ($t$ = 240 s in (a)), the magnetic field begins to deform (see panel (b)). Specifically, the gradient of the magnetic field near the null increases gradually from its initial value (see panel (b) and (c)) with an accompanying accumulation of electric current according to  Amp$\grave{\rm e}$re's law. The generation of the electric current gives rise to an unbalanced Lorentz force that reinforces the perturbation, and the null point locally collapses. It is noticeable that the Lorentz force is much stronger than the gradient in thermal pressure. The Lorentz force, shown by red arrows in the top-row of Figure~\ref{label 4a}, is directed in such a way that it initiates a gradual lateral thinning of the current accumulation region. Simultaneously, the unbalanced Lorentz force in the $y$-direction results in a vertical stretching of the current accumulated region (see top-row of Figure~\ref{label 4a}). This simultaneous thinning and stretching eventually gives rise to the formation of an elongated current sheet (see $t=541$~s, 660~s image-panels of the top-row of Figure~\ref{label 4a}; Figure4.mp4), where magnetic reconnection sets in later on. Thinning of the reconnection region CS occurs in opposite to the outward magnetic diffusion in the simulation. Gradual steepening of the impinging magnetic field with time as exhibited in panel (c) of Figure~\ref{label 3a}) also describes the reverse process of the magnetic field annihilation, i.e., it is the direct measure of the current sheet formation and its subsequent thinning. 

\begin{figure*}
\hspace{-1.5 cm}
\includegraphics[scale=0.28]{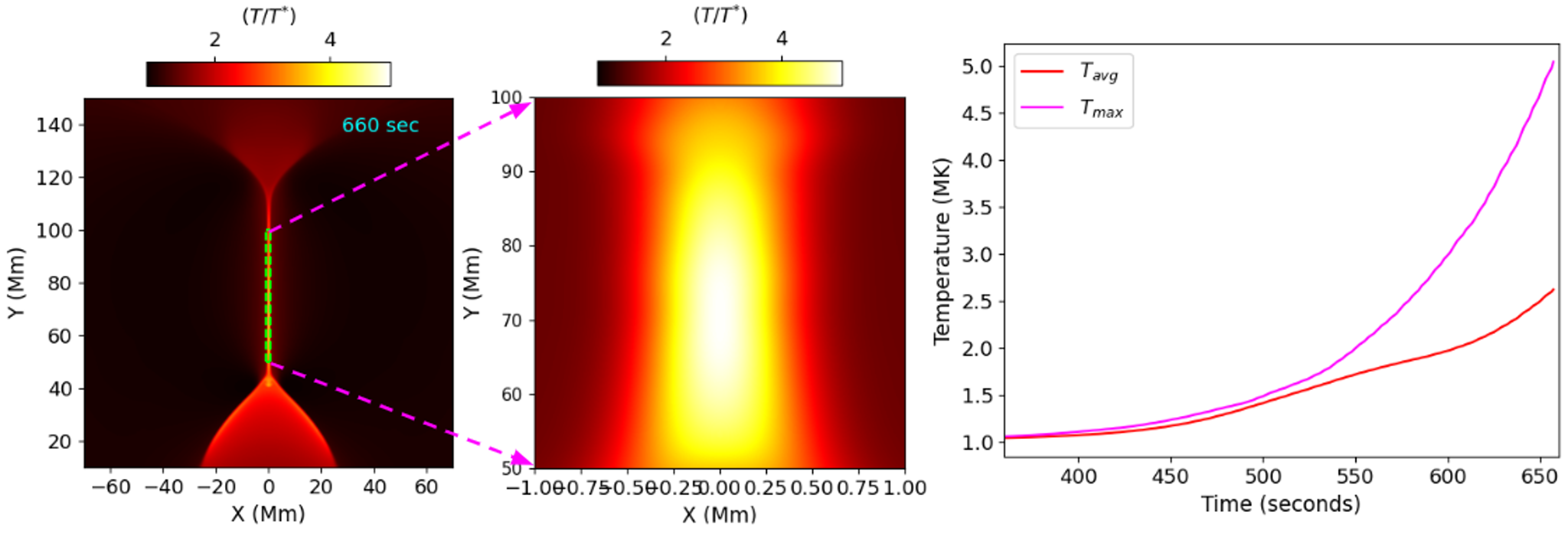}
\caption{Localized Heating of the Plasma: (a) A zoomed-in view of the reconnection region and current sheet in the numerical model and its extremely high temperature (top-left panel) are shown. The top-right panel demonstrates the temporal evolution of the average and maximum temperature along the current sheet.}
\label{label 5a}
\end{figure*}

\begin{figure*}
\hspace{-0.25 cm}
\includegraphics[height=22 cm, width=0.9\textwidth]{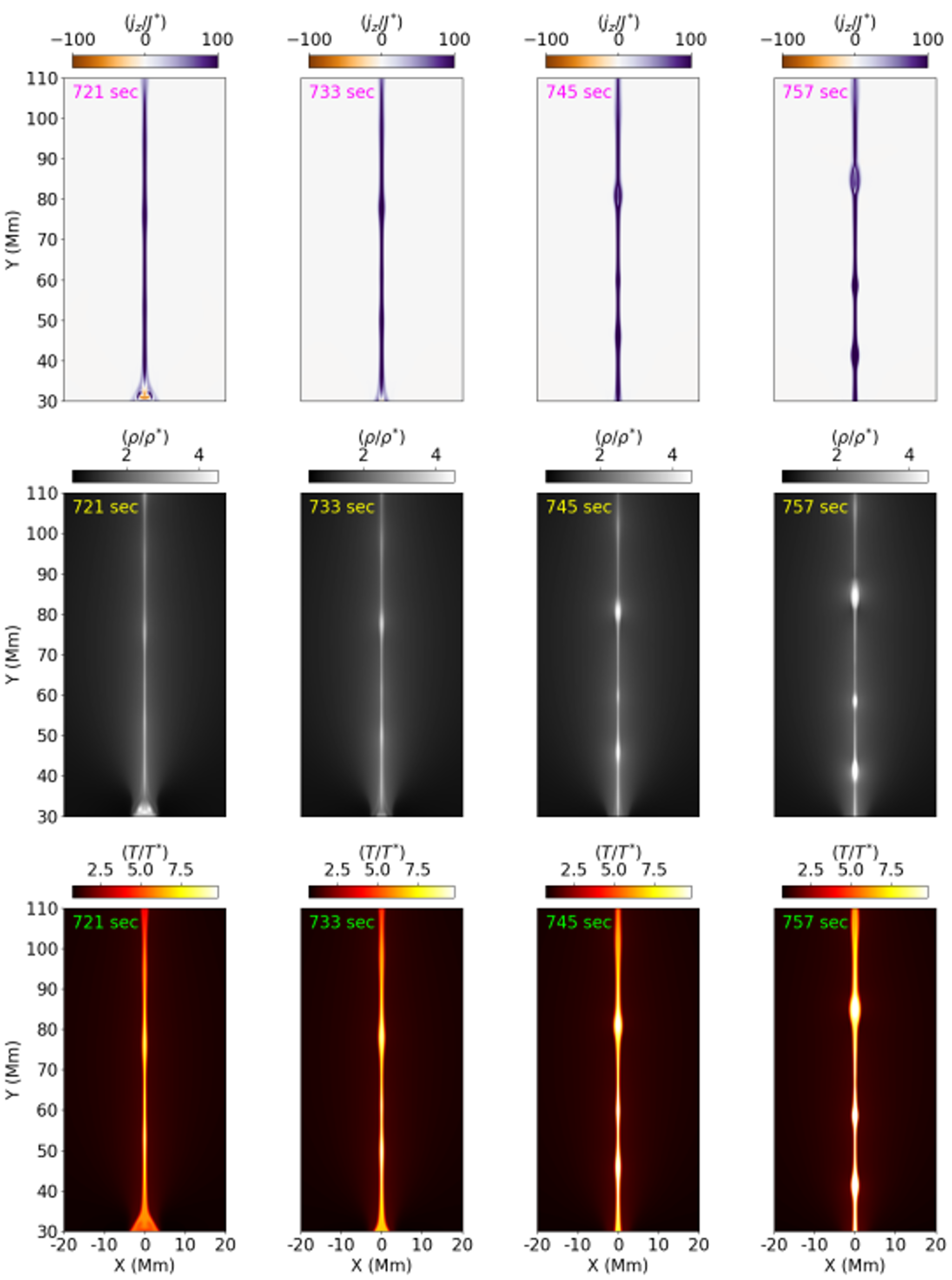}
\caption{Top: Magnetic $O$-type structures are shown along the elongated current sheet in the current density maps at four different times at a cadence of 12 s. Middle: Highly dense plasma blobs are depicted at the positions of the magnetic $O$-type structures as seen in the current density maps. Hence, the plasma blobs are basically plasmoids. Bottom: High temperature plasma is seen at the positions of these highly dense plasma blobs suggesting that these are reconnection generated plasmoids rather than entropy modes. The entire dynamics within the CS from 663 s to the formation and movement of plasmoid till 757 s is available as an animation Figure6.mp4 in the online version. The real-time duration of the animation is 4 s.}
\label {label 6a}
\end{figure*}

As we already noted above, the Lorentz force stretches and thins the magnetic field to
form the current sheet and triggers magnetic reconnection, the enhancement in the maximum of the current density with
time measured at Y = 92 Mm, i.e., the initial location of the magnetic null, indicates a gradual accumulation of current at that location from 469 s to 660 s (Figure~\ref{label 4a}, (b)). Figure~\ref{label 4a}, (c) also depcits the gradual thinning of the current sheet starting from 469 s to 660 s. At 469 s, the current density profile is fitted with a Gaussian profile to estimate the FWHM as a quantitative measure of the CS width, which subsequently reduces during the formation of CS. A similar process of forming a reconnection region has been found in other observed apparent current-sheet like structures recently reported in the on-disk and off-limb solar corona \citep{2024ApJ...974..104D}.  
Actually, the growth-rate (i.e., thinning of current-sheet) of the current-accumulated region is faster in the simulation than estimated for the observed magnetic structures in \citet{2024ApJ...974..104D}. There are several reasons why 
this is the case, including differences in the plasma-$\beta$, resistivity, magnetic structure,  projection effects, as well as different initiation mechanisms of thinning. 
In conclusion, the exponential profile of the modeled current-sheet thinning rates agree with recent statistical observations  \citet{2024ApJ...974..104D} and resembles  the concept of reconnecting coronal current sheets \citep{2024ApJ...963..139M}. In the present model, wave-driven collapse of a null region initiates the formation and gradual thinning of the CS (Figures~\ref{label 3a} \& ~\ref{label 4a}), which also subsequently undergoes reconnnection, heating, and fragmentation (Figures~\ref{label 5a} \& ~\ref{label 6a}).
The modeled  reconnection produces an increase in the temperature  within the current sheet (see Figure~\ref{label 5a}(a)). The maximum and average temperature of the current sheet rise are  $\approx$5.0 MK, and $\approx$2.5 MK, respectively. 
Moreover, since the temperature gradient builds up quickly in the reconnecting current sheet in the corona, we include thermal conduction which basically smooths out the temperature and keeps the current sheet evolving with an average temperature of a few mega-Kelvin.

Once the effective current sheet becomes simultaneously elongated and thinned enough to achieve fragmented structures along its length, it forms multiple plasmoids. The plasmoids become visible primarily around 720 seconds and then gradually grow in cross-sectional area with time. The bright plasma blobs seen in density maps (see middle panel of Figure~\ref{label 6a}) are similar to observed plasma blobs.
To demonstrate that these plasma blobs are plasmoids, we have plotted the current density to reveal a depleted current and O-type geometry, consistent with  previous simulations of plasmoid formation (see top panel of Figure~\ref{label 6a}). In the simulated dynamics, plasmoid formation starts approximately four minutes after the start of the current accumulation. In the bottom panel,  high-temperature plasma is seen at the positions of these highly dense plasma blobs suggesting that they are reconnection generated plasmoids. Such plasma blobs, in the effective presence of thermal conduction could have be linked with an entropy mode as previously depicted \citep[e.g.,][and references cited there]{2011A&A...533A..18M,2024MNRAS.528.5098G}. However, the physical properties of the dynamical plasma blobs  seen in Figure~\ref{label 6a} suggest that they are reconnection-generated plasmoids rather than entropy modes. 
The plasmoid formation depends on several factors such as: the rate of thinning and stretching of the current sheet which are in turn responses of the resistivity; the amplitude of the effective perturbations; and the compressibility of the plasma. The rapid rate of thinning  (see Figure~\ref{label 4a}) results in a rapid formation of plasmoids. However, the physics of their formation and evolution is essentially the same whether they are formed earlier or later, and it also depends on the plasma and magnetic field properties of the current sheet \citep{2025ApJ...979..207M}.

\section{Discussion and Conclusions}

Dissipation of MHD waves and magnetic reconnection have been widely studied as two potential mechanisms for sustaining high-temperature plasma in the solar corona, which is subjected to energy losses via thermal conduction, radiative cooling and solar winds. But, apart from a series of theoretical studies or numerical simulations \citep[e.g.,][]{2019ApJ...879..127T},
there has been little observational validation of an inter-linkage between MHD waves and magnetic reconnection. Rather, they have largely been studied  separately in the past, both numerically and observationally.  
Recently, \citet{2024NatCo..15.2667K} reported direct observation of mode conversion of MHD waves in the neighbourhood of a 3D magnetic null, but they did not mention reconnection or plasma heating. Thus, the co-existence of MHD waves and magnetic reconnection needs to be better explored, using simulations and observations in a detailed manner. Moreover, our understanding of the role of such processes in coronal heating and plasma dynamics has been hampered by an absence of direct observations. 

In the present paper, 
we demonstrate that when a fast magnetoacoustic wave passes through a region of the corona 
containing 
a magnetic null, it can trigger the formation of a reconnecting current sheet, followed by localized heating and plasma dynamics. We explore the physics of such a scenario via numerical simulation, 
which demonstrates the 
 accumulation of current in the magnetic null region, the gradual thinning and elongation of the  current sheet as it forms, and the later growth of plasmoids. 
In the simulation, the reconnecting current sheet is fully formed within 5 or 6 minutes of the passage of the fast magnetoacoustic perturbation front through the magnetic null. 
However, the entire suite of physical processes begins once these wave perturbations impinge on the null region and start to collapse it. Apart from the formation of the current sheet and the thinning, reconnection, heating, fragmentation and plasmoid formation are evident. 
Their co-existence and mutualism  depend on each other, which is an example of the Symbiosis of WAves and Reconnection (SWAR). 
Here, we find the generation of reconnection a favourable example for the process of SWAR. We are not considering the tiny reconnection scales generated at ion-gyro scales in the solar corona \citep[e.g.,][]{2010ApJ...720.1603G} or electron gyro-scales at Earth's magnetosphere \citep[e.g.,][]{2021GeoRL..48.0946Z}. It is clear that reconnection in the current sheet needs to take place to create the $O$-type magnetic field structures since they represent a change of field topology. Although the dissipation scale in the corona is not resolved in such a simulations, reconnection is facilitated by a spatially uniform resistivity that is larger than the Spitzer value, and ensures that the current structure remains resolved throughout the simulation (e.g., top-panel of Figure~\ref{label 6a}). Therefore, the bi-directional plasma outflows, heating, and plasmoid generations are clear outcomes of the reconnecting CS, and thus observable as bulk plasma processes  at MHD scales in the framework of $"$SWAR$"$.

Considering the importance of the aforementioned plasma processes in coronal heating, we note first that large-scale fast magnetic perturbations are often observed in the solar corona propagating across active regions and the quiet-Sun, for example, as EUV waves \citep[e.g.,][]{2011ApJ...736L..13L,2013ApJ...766...55K}. 
These waves can resonantly oscillate local coronal magnetic structures by partially transmitting energy and generating secondary waves within them \citep[e.g.,][]{
2012ApJ...753...52L,
2016MNRAS.463.1409S}. MHD waves and shocks may in principle carry enough energy to heat the solar corona, but it is uncertain whether they can dissipate energy efficiently and what is their filling factor in the  corona \citep[e.g.,][]{2020SSRv..216..140V}. For example, it has also been found  that adiabatic compression at the  front of fast-mode waves may heat the coronal plasma very little  \citep[e.g.,][]{2015ApJ...812..173V}. The dissipation of currents via magnetic reconnection is another way in principle of heating the corona, but whether this is by braiding or by flux cancellation is unclear \citep[e.g.,][]{2024ApJ...960...51P}. 

The role of co-existing reconnection and waves in heating  the solar corona is little understood due to the lack of direct observations, but here we suggest there may be a $``$Symbiosis of WAves and Reconnection (SWAR)$"$, in which both play a part. 
The example we are presenting here is of waves from flares or microflares driving reconnection in the ambient corona. Almost half of the large solar flares  are observed to be accompanied by fast magnetoacoustic waves sweeping across the corona far from the flare sites.
Such waves are therefore likely to cause heating at the myriads of null points that are scattered throughout the chromosphere and corona \citep[e.g.,][]{2009SoPh..254...51L}.
The same effect will also occur at quasi-separators in the absence of nulls \citep[e.g.,][]{2022LRSP...19....1P}. Since flares occur on a wide range of scales and sizes and these time-dependent bursts of energy will always generate the fast-mode waves and shocks \citep[e.g.,][]{2009SoPh..254...51L}, it is to be expected that the large events that we have modeled here represent a process that is likely to be present also at much smaller scales throughout the corona and chromosphere. Furthermore, since the density of magnetic nulls is very much greater in the chromosphere than the corona \citep[e.g.,][]{2003PhPl...10.3321L}, wave-driven reconnection is likely to provide more heating of the chromosphere than the corona. In conclusion, in this paper,  a numerical example of  wave-reconnection symbiosis is presented as a potential mechanism for heating and dynamics of the solar atmosphere.

\section*{acknowledgments}
AKS acknowledges the ISRO grant (DS/2B-13012(2)/26/2022-Sec.2) for the support of his scientific research. SM thankfully acknowledge the Prime Minister Research Fellowship (PMRF) to support his research. We acknowledge open source framework of AMR-VAC. D.I.P. gratefully acknowledges support through an Australian Research Council Discovery Project (DP210100709). This study was supported by the National Natural Science Foundation of China (NSFC,12173012,12473050), Guangdong Natural Science Funds for Distinguished Young Scholars (2023B1515020049) and Shenzhen Science and Technology Project (JCYJ20240813104805008). SKM acknowledges the Kyoto University, Japan for postdoctoral research grant to him. KM's work was done within the framework of the project from the Polish Science Center (NCN) Grant No. 2020/37/B/ST9/00184. Authors acknowledge the scientific discussions with Prof. Leon Ofman (NASA-GSFC).

\facilities{SDO/AIA}
\vspace{5mm}
\software{MPI-AMRVAC, Paraview, Python}

\end{document}